\def \R {{\it R}}
\def \d {\hbox{d}\,}
\def \square {\hbox{$\sqcup\!\!\!\!\sqcap$}}
\def \s {\square^{-1}}
\def \sgn {\hbox{sgn}}
\def \e {\hbox{e}}

\def \v {\vskip}

\def \be {\begin{equation}}
\def \ee {\end{equation}}
\documentstyle[12pt]{article}

\begin{document}


The interest of studying lower-dimensional gravity has been growing
in the last years.   Although these  theories  have  their   own
physical significance, they are also studied as toy models for the
$3+1$ dimensional theory.

In  $2+1$ dimensions,  and in the absence  of  matter,  Einstein's
general relativity can be formulated as a Chern-Simons theory with
an $ISO(2,1)$ gauge group \cite{[1]}.  In the new formulation  the
quantum  Hilbert  space of the theory can  be  determined  exactly
although most of the conceptual issues of quantum gravity  remain
to be addressed.  In $1+1$ dimensions, the Hilbert-Einstein action
does not have physical content since it is a topological invariant
of   the   space-time  manifold  (the   Einstein tensor   vanishes
identically). In Ref.\cite{[2]}, the equation $R=\frac{\Lambda}{2}$
was  suggested  as  the  natural analog  of  the  vacuum  Einstein
equations   with  a cosmological constant.  This equation  can  be
enforced from a local variational principle if a scalar  (dilaton)
field $\Phi$ is present in the theory \cite{[2]}:

\begin{eqnarray}
S = \int\sqrt{g}\left(R - \frac{1}{2}\Lambda\right)\Phi\>.\label{1}
\end{eqnarray}
The  requirement  of constant curvature (Liouville  equation) can
also be derived from the action induced by massless matter  fields
\cite{[3]}, i.e., the so-called induced 2d-gravity:

\begin{eqnarray}S = \frac{c}{96\pi}\int\,\sqrt{g} (R\s R +
\Lambda)\label{2}\end{eqnarray}
This action can also be written as a local one if we introduce a
(dilaton) scalar field $\Phi\equiv\s R$ (we shall omit the
coupling constant)

\begin{eqnarray}\frac{1}{2}\int\sqrt{g}(g^{\mu\nu}
\partial_\mu\Phi\partial_\nu\Phi +
2R\Phi + \Lambda)\;.\label{3}\end{eqnarray}

On  the other hand,  in $3+1$ dimensions the attempts to  quantize
the  gravitational  field  have a long history.  One  of  the  most
interesting   approaches  to  apply  quantum  mechanics   to   the
gravitational   field  is   the   so-called   Quantum   Cosmology
\cite{[4],[5],[6]}.  The  idea  is to restrict the metric  and  matter
fields of the theory to a finite number of degrees of freedom  and
then   quantize   the   reduced   theory (generally   known   as
"minisuperspace" model). In this approach, and in canonical
quantum gravity in general, the quantum wave functions must
satisfy the Wheeler-DeWitt equation, i.e., the analogue of the
Schr\"odinger equation for generally covariant theories.
Related problems are the issue of time, the
probability measure for the scalar product and how to single out
an unique quantum state representing the "ground state" of the
theory (see the reviews \cite {[7],[8]}).

In this letter we shall consider the induced 2d-gravity  (\ref{3})
on  a  closed spatial section in the  minisuperspace  (or  quantum
cosmology)  approach.  We  shall give the general solution  to  the
corresponding Wheeler-DeWitt (WdW) equation providing the proper Hilbert
space of the model. Finally we shall discuss the probability distribution of
the elementary solutions and also the validity of  the
minisuperspace approximation.

We construct now the minisuperspace model of the theory  (\ref{3})
by  restricting  the metric and the scalar field (dilaton)  to  be
spatially homogeneous:
\begin{eqnarray}
\d s^2 &=& N^2(t)\d t^2 - a^2(t)\d \theta^2\\ \label{4}
\Phi &=& \Phi(t)\>,\label{4a}
\end{eqnarray}
where $N=N(t)$ is the lapse function, $a=a(t)$ is the radius of
the 2d universe, and the shift function has been set to zero: $N^i
\equiv0$. In terms of the variables $N$, $a$ and $\Phi$ the action
(\ref{3}) becames
\be\int\,\d t\left[\frac{1}{2}\frac{a}{N}\dot\Phi^2 +
2\frac{\dot{a}\dot\Phi}{N} +
\frac{\Lambda}{2}Na\right]\>.\label{5}\ee
This action takes the following hamiltonian form (the lapse
function $N$ plays the role of a lagrange multiplier):

\be\int\,\d t\left[\pi_\Phi\dot\phi + \pi_a\dot a -
N{\cal C}\right]\>,\label{6}\ee
where the momenta $\pi_\Phi$ and $\pi_a$ are

\begin{eqnarray}
\pi_\Phi&=&\frac{a}{N}\dot\Phi + 2\frac{\dot a}{N}\label{7}\\
\pi_a&=&\frac{2}{N}\dot\Phi\>,\label{8}
\end{eqnarray}
and the hamiltonian constraint ${\cal C}$ is
\be {\cal C} = -\frac{1}{8}a\pi_a^2 + \frac{1}{2}\pi_a\pi_\Phi -
\frac{\Lambda}{2}a\label{9}\ee

The classical constraint ${\cal C}= 0$ is enforced at the quantum
level by imposing that the wave functions are annihilated by the
operator version of (\ref{9}), i.e., the Wheeler-DeWitt equation.
As usual we face then the standard factor ordering ambiguities. In
our case we shall fix the operator ordering of the classical
function (\ref{9}) as follows:

\begin{eqnarray}
\Bigl[&-&\frac{1}{8}
\left(-i\hbar a\frac{\partial}{\partial a}\right)\frac{1}{a}
\left(-i\hbar a\frac{\partial}{\partial a}\right) \nonumber \\&+&
\frac{1}{4}\left[\frac{1}{a}\left(-i\hbar a\frac{\partial}{\partial a}\right)+
\left(-i\hbar a\frac{\partial}{\partial
a}\right)\frac{1}{a}\right]\left(-i\hbar\frac{\partial}{\partial
\Phi}\right)\nonumber  \\
&-&\frac{\Lambda}{2}a\Bigr]\Psi(\Phi,a) =
0\>.\label{10}\end{eqnarray}

We can justify this choice by arguing that, due to the fact that
the classical variable $a$ satisfies the inequality $a>0$, the
operator $\hat \pi_a=-i\hbar\frac{\partial}{\partial a}$ fails to be
self-adjoint in $L^2(\R^+,\d a)$. The natural self-adjoint
operators on the positive real line, respect to the measure
$\frac{\d a}{a}$, correspond to the  variables $a, p_a(\sim
a\pi_a)$ \cite{[9]}, which verify the affine commutation relation
$[\hat a, \hat p_a] = i\hbar\hat a$. As is well known, the unitary
representation on $L^2(\R^+,\frac{\d a}{a})$ of the affine group
$\hat a = a, \hat p_a = -i\hbar a\frac{\partial}{\partial a}$
preserve the positivity of the $\hat a$-operator spectrum. In
terms of the affine variables, the constraint ${\cal C} = 0$
becomes
\be -\frac{1}{8}\frac{1}{a}p_a^2
+\frac{1}{2}\>\frac{1}{a}p_a\pi_\Phi - \frac{\Lambda}{2}a = 0\>.
\label{11}\ee
The quantum version of (\ref{11}) leading to an hermitian Wheeler-
DeWitt operator $\hat{\cal C}$ is just (\ref{10}).

To solve the Wheeler-DeWitt equation (\ref{10}) we should  realize
that   the  operator  $\hat\pi_\Phi$  commute  with  the   quantum
constraint  (in fact,  $\pi_\Phi$ is a constant  of  motion).  This
implies  that  the Wheeler-DeWitt equation separates  and  we  can
expand the general solution in $\hat\pi_\Phi$-eigenstates:

\be\Psi(\Phi,a)= \int\,\d \lambda\e^{\frac{i\lambda\Phi}{\hbar}}
\Psi_\lambda(a)\>,\label{12}\ee
where the functions $\Psi_\lambda(a)$ satisfy

\be\left(a^2\frac{\d^2}{\d a^2}-
\frac{4i\lambda}{\hbar}a\frac{\d}{\d a} + \frac{2i\lambda}{\hbar}-
\frac{4\Lambda}{\hbar^2}a^2\right)\,\Psi_\lambda(a) =
0\>.\label{13}\ee

If we make now a change of variable,
$x=\frac{2\sqrt{|\Lambda|}}{\hbar}a$, and function,
$\Psi_\lambda\equiv x^{\frac{1}{2}+\frac{2i\lambda}{\hbar}}u_\lambda$,
we arrive at

\be x^2 u''_\lambda + xu'_\lambda + \left(-\sgn\Lambda x^2 -
\nu^2\right)\,u_\lambda = 0, \label{14}\ee
where
\be \nu^2 = \frac{1}{4}\left(1-\frac{16\lambda^2}{\hbar^2}\right).
\label{15}\ee
For $\Lambda<0$, (\ref{14}) is a Bessel equation whose general
solution is an arbitrary lineal combination of Bessel and Neumann
functions J$_\nu$ and N$_\nu$. If $\Lambda >0$, the solutions are
lineal combinations of the modified Bessel and Hankel functions
I$_\nu$ and K$_\nu$.

Once we know the general solution of the Wheeler-DeWitt equation
of our minisuperspace model we have to construct a probability
measure, i.e., a scalar product, to give the proper physical
Hilbert space of the model. The scalar product measure for the
minisuperspace functions $\Psi(a,\Phi)$ respect to which the
Wheeler-DeWitt operator was made self-adjoint was just
$\frac{\d a}{a}\d \Phi$.
It is therefore natural to propose this measure to
construct an inner product for the solutions of the Wheeler-DeWitt
equation. However, this proposal has the general drawback that
one of the minisuperspace configuration coordinate plays the role
of "time" and using the measure $\frac{\d a}{a}\d \Phi$ could
mean we are integrating over the time coordinate as well as the
proper configuration variable (see for instance \cite{[8]}). We
should have in mind this fact to interpret the resulting
expression of the scalar product.

Given a general solution of the Wheeler-DeWitt equation for a
negative cosmological constant we can expand it as (Re $\nu\ge0$,
Im $\nu\ge0$):

\be \Psi = \int\,\d \lambda \e^{i\lambda\Phi/\hbar}x^{\frac{1}{2}
+ \frac{2i\lambda}{\hbar}}\left( A(\lambda)\hbox{J}_\nu(x) +
B(\lambda)\hbox{N}_\nu(x)\right)\,, \label{16}\ee
where $A(\lambda)$ and $B(\lambda)$ are two arbitrary complex
functions and the norm of the solutions (\ref{16}) (with the meaure
$\frac{\d a}{a}\d \Phi$) becomes

\be <\Psi|\Psi> = \int_0^\infty\d
x\int\,\d\lambda\left[|A|^2|\hbox{J}_\nu|^2 +
|B|^2|\hbox{N}_\nu|^2 + A^*B\hbox{J}_\nu^*\hbox{N}_\nu +
 AB^*\hbox{J}_\nu\hbox{N}^*_\nu\right].\label{17}\ee
But now the above integrals respect to the
$x$-variable are divergent. This fact could be understood
as the effect of the integration respect to some sort
of "time" variable. However we can define, in a natural way, a
regularized scalar product by substituting $\d x$ in (\ref{17})by
$\d x/x^\epsilon$. All the integrals respect to the $x$-
variable are then proportional to the common factor
$\Gamma(\epsilon)/2^\epsilon$. We shall absorbe this overall factor
and take the limit $\epsilon\rightarrow 0$ to define the physical
scalar product. For instance, when $B=0$ the regularised scalar
product turns out to be

\be <\Psi|\Psi>_{_{\hbox{reg}}}= \int\d\lambda\frac{|A|^2}
{\Gamma(\frac{\nu-\nu^*+1}{2})\Gamma(\frac{\nu^*-\nu+1}{2})}\>.\label{18}
\ee

Let us study now the probability
distribution $x^{-1}|\Psi|^2$
of the elementary solutions according to the possible values
(\ref{15}) of the order $\nu$:

$\>1$a) If Im $\nu=0$ and Re $\nu \in ]0,1/2]$ the probability
distribution (up to normalization) $|\hbox{J}_\nu|^2$ is damped for
$x<1$, vanishes at the origin $x=0$, and oscillates for $x>1$
decaying slowly for $x\rightarrow +\infty$
\begin{eqnarray}&\hbox{J}_\nu\sim_{_{\!\!\!\!\!\!\!\!\!{x\rightarrow0}}}
\frac{1}{\nu!}
\left(\frac{x}{2}\right)^\nu\label{19}\\
&\hbox{J}_\nu\sim_{_{\!\!\!\!\!\!\!\!\!\!\!{x\rightarrow
+\infty}}}\left(\frac{2}{\pi
x}\right)^{\frac{1}{2}}\cos[x-\frac{\pi}{2}(\nu+\frac{1}{2})]\>.
\label{20}\end{eqnarray}
Note that the maximum amplitude for the
radius is obtained for $\nu=1/2$ (i.e., when $\lambda=0$).

$\>1$b) If Im $\nu=0$, and Re $\nu\in]-1/2,0[$ the probability
distribution $|\hbox{N}_\nu|^2$ is now peaked around the origin $x=0$
$(|\hbox{N}_\nu|^2\sim_{_{\!\!\!\!\!\!\!\!\!{x\rightarrow0}}}
\hbox{cotg}(\nu\pi)\frac{1}{\nu!}
\left(\frac{x}{2}\right)^\nu)$.

$\>2$a) For $\nu=0$, $|\hbox{J}_0|^2$ oscillates for $x>1$ decaying
slowly for $x\rightarrow+\infty$ as the case $1$a) but it
does not vanish at the origin.

$\>2$b) For $\nu=0$ the general behaviour of $|\hbox{N}_0|^2$ is
similar to $1$b)but the asymptotic expression for $x\sim 0$ is now
logaritmic
\be \hbox{N}_0\sim_{_{\!\!\!\!\!\!\!\!\!{x\rightarrow0}}}
\frac{2}{\pi}\hbox{ln}
\left(\frac{\gamma x}{2}\right)\>.\label{21}\ee

$\>3$a,b) If Re $\nu = 0$, we have oscillatory wave functions,
$\hbox{J}_\nu$ and N$_\nu$, decaying slowly for $x\rightarrow \infty$,
but with a non-vanishing amplitude at $x=0$.

On the other hand, when the cosmological constant is positive,
only the modified Hankel functions $\hbox{K}_\nu$ lead to normalizable
quantum wave functions. According to the value of $\nu$
we have the following possibilities:

$\>1$) If Im $\nu=0$, the probability distribution is peaked
around the origin
\be \hbox{K}_\nu \sim_{_{\!\!\!\!\!\!\!\!\!{x\rightarrow0}}}
\frac{\pi}{2\sin\pi|\nu|}
\frac{1}{-|\nu|!}\left(\frac{2}{x}\right)^{|\nu|}\>,\label{22}\ee
decaying exponentially for large $x$

\be \hbox{K}_\nu
\sim_{_{\!\!\!\!\!\!\!\!\!{x\rightarrow0}}}
\left(\frac{\pi}{2x}\right)^{1/2}\e^{-x}\>.\label{23}\ee
Note that for $\lambda = 0$ neither the solutions $\nu = 1/2$ or
$\nu = -1/2$ are normalizable.

$\>2$) If $\nu = 0$ the probability distribution is still
concentrated at the origin but their asymptotic behaviour is now
logaritmic
$\left(\hbox{K}_0\sim-\hbox{ln}\left(\frac{\gamma x}{2}\right)\right)$.

$\>3$) For Re $\nu=0$, we have oscillatory wave functions around
$x\sim 0$ but still decaying exponentially (\ref{23})
for $x\rightarrow +\infty$.

It is clear from the above analysis that a kind of phase transition
happens at $\nu=0$. For negative cosmological constant,
the critical point $\nu =0$ separates three different phases: the
damped phase $1$a), the collapsed phase $1$b) and the oscillatory
phase $3$a,b). For $\Lambda>0$, we have instead two different
phases: the collapses phase $1$) and the oscillatory phase $3$).

It is interesting to note that the different behaviour of the wave
function for large $x$ reflects appropiately the properties of the
classical solutions. In the conformal gauge, $a(t)=|\Lambda|^{-
1/2}N(t)$, the classical metric solutions are (we have chosen units
in which $\hbar=1=c$)

\be a(t) = \frac{8}{|\Lambda|^{\frac{3}{2}}}
\frac{p^2\e^{2pt}}{\left(1+\sgn\Lambda\e^{2pt}\right)^2}\>,\label{24}
\ee
where $p$ is an arbitrary real parameter. If $\Lambda >0$ there is
a maximum radius $\frac{8p^2}{|\Lambda|^\frac{3}{2}}$ reflecting
the exponential decay of the wave function for
$x\rightarrow+\infty$, whereas for $\Lambda<0$ the classical
radius is unbounded in agreement with the slow decay of the wave
functions. Moreover, the classical momenta $\pi_\Phi$ is
proportional to the "$p$" parameter and then we can easily understand the
absence of normalizable solutions when
$\lambda=0$, if $\Lambda > 0$, as the absence of classical
parabolic solutions.
For $\Lambda <0$, the parabolic solution does exist and,
coinciding with the limit  $p\rightarrow 0$ of (\ref{24}),
it is given by

\be a(t)= \frac{2}{|\Lambda|^\frac{3}{2}}\frac{1}{t^2}\>.\label{25}
\ee

Finally we would like to comment briefly on the validity of the
minisuperspace treatment of the $2$d-induced gravity. It was showed
in Ref. \cite{[10]} that the general solution for the metric and
the scalar field of the induced gravity admits a spatially
homogeneous expression. The solutions for the metric are
(\ref{24},\ref{25}) and the corresponding ones for the scalar field are

\be \Phi =\hbox{ln} \alpha \left(1
+\sgn\Lambda\e^{2pt}\right)^2\>,\label{26}\ee
where $\alpha$ is an arbitrary positive constant. We remark that
only the hyperbolic and parabolic monodromies lead to
classical solutions for the scalar field. According to this we can conclude
that the minisuperspace approximation to the induced gravity does not really
imply a restriction on the physical degrees of freedom. In fact, the
reduced phase space of the theory is finite dimensional and we
could argue, therefore, that we have provided, in some sense, an
exact description of the Hilbert space of the induced 2d-gravity.
We believe that the question of the ground state of the theory
deserves a separate study.

To finish we would like to point out that a similar treatment can
be carried out for the recently introduced "string-inspired" model
describing two-dimensional black hole physics \cite{[11]} (see
also \cite{[12]}).

\v 1cm

\noindent {\bf Acknowledgements.} M. N. is grateful to the CSIC and
the MEC for a FPI grant. M.N. acknowledges the Imperial College
for its hospitality.

\vfil
\eject

\vfil
\eject


\title{WAVE FUNCTIONS OF THE INDUCED 2D-GRAVITY\thanks{
Work partially supported by the Comisi\'on Interministerial de
Ciencia y Tecnolog\'\i a.}}

\author{Jos\'e Navarro-Salas$^{1,2}$, Miguel Navarro$^{2,3,4}$
and V\'\i ctor Aldaya$^{2,4}$}
\maketitle

\v 1.5cm

\noindent 1.- Departamento  de  F\'\i sica  Te\'orica, Burjassot-46100,
Valencia, Spain.
\v 0.3cm
\noindent 2.- IFIC, Centro Mixto Universidad de
Valencia-CSIC, Burjasot 46100-Valencia, Spain.
\v 0.3cm
\noindent 3. The Blackett Laboratory, Imperial College, London SW7 2BZ;
United Kingdom.
\v 0.3cm
\noindent 4.- Departamento  de  F\'\i sica  Te\'orica  y  del  Cosmos,
Facultad  de  Ciencias, Universidad de Granada, Campus de Fuentenueva,
Granada 18002, Spain.

\begin{abstract}

We consider the induced 2d-gravity in the minisuperspace approach.
The general solution to the Wheeler-DeWitt equation is given in terms
of different kind of Bessel functions of purely real or imaginary orders.
We study the properties of the corresponding probability distributions finding
a kind of phase transition at the critical point $\nu =0$.
\end{abstract}


\begin{thebibliography}{99}

\v 0.5cm
\bibitem{[1]}E. Witten, {\it Nucl. Phys.} {\bf B311} (1988)46.


\bibitem{[2]}R. Jackiw, in: {\it Quantum Theory of Gravity}, ed. S.
Christensen (Adam Hilger, Bristol, 1984)p. 403; C. Teitelboim, in
{\it Quantum Theory of Gravity}, ed. S. Christensen (Adam Hilger,
Bristol, 1984)p. 327.

\bibitem{[3]}A.M. Polyakov, {\it Mod. Phys. Lett.} {\bf A2} (1987) 899;
V.G. Knizhnik, A.M. Polyakov and A.B. Zamolodchikov, {\it Mod. Phys. Lett}
{\bf A3} (1989)819.

\bibitem{[4]}B.S. DeWitt, {\it Phys. Rev.} {\bf 160} (1967)1113;
C.W. Misner, {\it Phys. Rev.} {\bf 186} (1969)1319.

\bibitem{[5]}J.B. Hartle and S.W. Hawking, {\it Phys. Rev.} {\bf D28}
(1983)2960.

\bibitem{[6]}A. Vilenkin, {\it Phys. Rev.} {\bf D37} (1988)888.

\bibitem{[7]}J.J. Halliwell, in {\it Proceedings of the Jerusalem Winter
School on Quantum Cosmology and Baby Universes},ed. T. Piran (1990).

\bibitem{[8]}C.J. Isham, Imperial/{\bf TP}/90-91/14.


\bibitem{[9]}J.R. Klauder and E. Aslaksen, {it Phys. Rev.}
{\bf D}2 (1970)393; C.J.Isham, in: {\it Relativity, Groups and
Topology} II, Proc. 1983. {\bf Les Houches Summer School}, ed. B.
DeWitt and R. Stora (Amsterdam:North-Holland);
J. Navarro-Salas and J.R. Klauder, {\it Class. Quantum. Grav.}
{\bf 7} (1990)1207.

\bibitem{[10]}J. Navarro-Salas, M. Navarro and V. Aldaya,CERN-TH.6537/92.
Revised version FTUV-92-56. See also {\it Phys. Lett.} {\bf B292} (1992)19.

\bibitem{[11]}C.G. Callan, S.B. Giddings, J.A. Harvey and A. Strominger,
{\it Phys. Rev.} {\bf D45} (1992); H. Verlinde. in {\it The Sixth Marcel
Grossman Meeting on General Relativity}, ed. H. Sato (World Scientif.
Singapore, 1992).

\bibitem{[12]}R. Jackiw, MIT preprint {\bf CTP}\#2125 (July 1992).


\end{thebibliography}
\end{document}